\documentclass[10pt,pre,onecolumn,superscriptaddress,reprint]{revtex4-2}
\usepackage{xr}
\usepackage{graphicx}
\usepackage{float}
\usepackage{amsmath}
\usepackage{amssymb}
\usepackage{mathtools} 
\usepackage{color}
\usepackage{xcolor}
\usepackage[english]{babel}
\usepackage{silence}
\usepackage[T1]{fontenc}
\usepackage[utf8]{inputenc}

\usepackage[columnwise]{lineno}

\usepackage{hyperref}
\usepackage{physics}
\usepackage[caption=false]{subfig}

\begin{document}

\title{Ordered and disordered stealthy hyperuniform point patterns across spatial dimensions}

\date{\today}
\author{Peter K. Morse}
\thanks{Corresponding author.}
\email{peter.k.morse@gmail.com}
\affiliation{Department of Chemistry, Princeton University, Princeton, NJ 08544}
\affiliation{Department of Physics, Princeton University, Princeton, NJ 08544}
\affiliation{Princeton Institute of Materials, Princeton University, Princeton, NJ 08544}
\author{Paul J. Steinhardt}
\affiliation{Department of Physics, Princeton University, Princeton, NJ 08544}
\author{Salvatore Torquato}
\thanks{Corresponding author.}
\email{torquato@princeton.edu}
\affiliation{Department of Chemistry, Princeton University, Princeton, NJ 08544}
\affiliation{Department of Physics, Princeton University, Princeton, NJ 08544}
\affiliation{Princeton Center for Theoretical Science, Princeton University, Princeton, NJ 08544}
\affiliation{Princeton Institute of Materials, Princeton University, Princeton, NJ 08544}
\affiliation{Program in Applied and Computational Mathematics, Princeton University, Princeton, NJ 08544}

\begin{abstract}
In previous work [Phys. Rev. X \textbf{5}, 021020 (2015)], it was shown that stealthy hyperuniform systems can be regarded as hard spheres in Fourier-space in the sense that the the structure factor is exactly zero in a spherical region around the origin in analogy with the pair-correlation function of real-space hard spheres. In this work, we exploit this correspondence to confirm that the densest Fourier-space hard-sphere system is that of a Bravais lattice. This is in contrast to real-space hard-spheres, whose densest configuration is conjectured to be disordered. We also extend the virial series previously suggested for disordered stealthy hyperuniform systems to higher dimensions in order to predict spatial decorrelation as function of dimension. This prediction is then borne out by numerical simulations of disordered stealthy hyperuniform ground states in dimensions $d=2$-$8$.
\end{abstract}

\maketitle

\section{Introduction}
Hyperuniform systems are defined by a structure factor $S(\mathbf{k})$ which approaches zero as the wavenumber $k\equiv\abs{\mathbf{k}}$ approaches zero, yielding density fluctuations that are anomalously suppressed at long length scales as compared to standard liquids or gases~\cite{torquato_local_2003}. These systems describe states that are ubiquitous in nature ranging from ordered systems, like crystals and quasicrystals, to disordered systems like perfect glasses, fermionic point processes, jammed particle packings, quantum states, plasmas, galaxy distributions, and eigenvalues of random matrices~\cite{torquato_hyperuniform_2018}. Hyperuniform systems can be classified according to the precise way the structure facgtor approaches zero, which is useful because different classes have distinctive properties and applications.

\begin{figure}[bp]
\includegraphics[width=0.95\columnwidth]{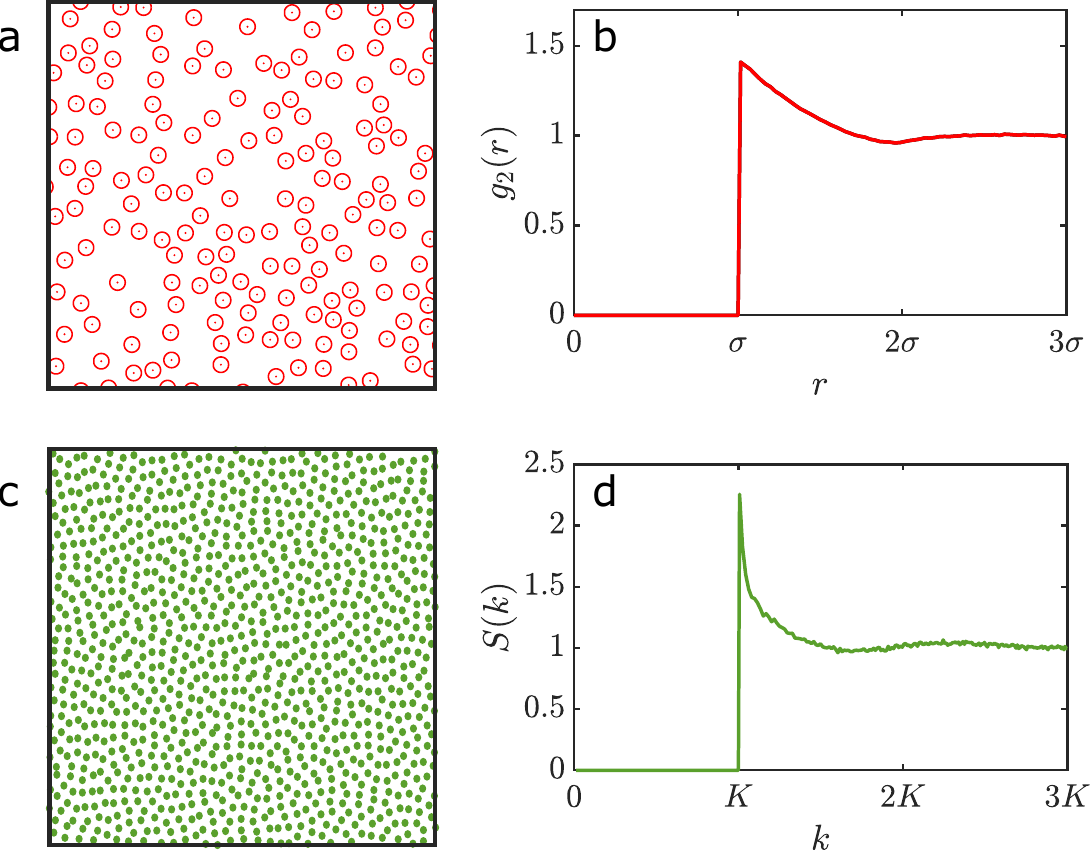}
\caption{A comparison between (a) real-space hard-disks with (b) their exclusion region in $g_2(r)$ and (c) a disordered stealthy hyperuniform point pattern (Fourier-space hard-spheres) with (d) their exclusion region in $S(k)$.}
\label{fig:diagram}
\end{figure}

Stealthy hyperuniform point-patterns are a sub-class of hyperuniform states wherein $S(k)$ is precisely zero within an exclusion region $0 < k \leq K$ for some positive $K$. This property suggests an analogy: up to two particle correlations, stealthy hyperuniform point patterns can be regarded as hard spheres in Fourier-space~\cite{torquato_ensemble_2015} in the sense that the pair correlation function for hard spheres, $g_2(r)$, has an exclusion region in real space, where $g_2(r)$ is precisely zero for $0<r\leq\sigma$, where $r$ is the distance between particles and $\sigma$ is the sphere diameter (see Fig.~\ref{fig:diagram} for a comparison). The analogy breaks down when considering higher order correlations ($g_n$ with $n>2$), but the correspondence is strong enough to deduce several properties of stealthy hyperuniform systems. While bearing in mind this qualification, we use the terms stealthy hyperuniform point pattern and Fourier-space hard-spheres interchangeably.

In considering the analogy, it is natural to extend the concept of packing fraction for hard spheres in real space to Fourier space, as in Ref.~\cite{torquato_ensemble_2015}. For hard-spheres in real space, the packing fraction is proportional to $\sigma^d$, where $d$ is the dimension of space. In Fourier space, the efective packing fraction of stealthy hyperuniform configurations is proportional to $K^d$~\cite{torquato_ensemble_2015}. In the special case of crystalline stealthy hyperuniform systems, $K$ is simply the location of the first Bragg peak~\cite{uche_constraints_2004}, while for disordered stealthy hyperuniform systems, $K$ is the smallest wavenumber for which $S(K)$ is positive. This suggests a simple question: is there a direct correspondence between ideal hard-sphere packings in real-space and in Fourier-space (i.e. solutions to the densest packing, the covering problem, or the quantizer problem~\cite{conway_sphere_1993, torquato_reformulation_2010})? 

In real-space, each of these problems is quite difficult. In particular, the densest packing has only been rigorously proven in dimensions $d=1$ (a trivial solution), $d=2$~\cite{lagrange_recherches_1773}, $d=3$~\cite{hales_proof_2005}, $d=8$~\cite{viazovska_sphere_2017}, and $d=24$~\cite{cohn_sphere_2017}. For all other dimensions, only lower~\cite{venkatesh_note_2013} and upper bounds~\cite{kabatiansky_bounds_1978, cohn_sphere_2014} have been proven, though conjectured but unproven densest packing candidates are often quoted for $d \le 48$~\cite{conway_sphere_1993}. Most of these candidates are Bravais lattices, though notable exceptions exist in $d=10$-$13$~\cite{best_binary_1980}, where the densest known packings are $n$-component crystals derived through binary codes. The question of the densest crystalline hard-sphere packing in real-space is thus made more difficult by the existence of infinitely many non-Bravais lattices (such as the honeycomb lattice in $d=2$ and hexagonal close packing in $d=3$). To complicate matters further, it has been conjectured that in large enough $d$, the densest packing should be disordered due to the decorrelation principle~\cite{torquato_new_2006}, which states that as $d\to\infty$ all unconstrained spatial correlations vanish beyond the hard-core and all $n$-particle correlation functions $g_n$ for $n\ge3$ can be inferred from $g_2$ and the number density. However, two other possibilities have been suggested through thermodynamic arguments as discussed in Ref.~\cite{charbonneau_three_2021}.

For Fourier-space hard spheres, the densest packing in any dimension has constraints that do not exist in real-space, allowing for stronger restrictions on its properties. In $d=1$-$4$, it was proven~\cite{torquato_ensemble_2015} that the densest Fourier-space hard-sphere packing in is a Bravais lattice---specifically the dual to the densest Bravais lattice in real-space. This statement was then conjectured to hold for higher dimensions---which we address in Sec.~\ref{sec:densest}. The densest configuration in Fourier-space thus does not necessarily correspond to the densest configuration in real-space except in cases where the densest lattice is self-dual. Here we derive an expression that makes this statement explicit, thus allowing the full phase diagram of stealthy systems to be computed.

We also comment directly on the decorrelation principle as applied to disordered stealthy hyperuniform states. In Ref.~\cite{morse_generating_2023}, we demonstrated the ability to create large disordered stealthy hyperuniform systems to ultra high accuracy by adapting their standard generating protocol to double-double precision on GPUs. With ultra-high accuracy systems at our disposal, it is then possible to test the decorrelation principle through $g_2(r)$, $S(k>K)$ and the $\tau$ order metric (which measures the degree of translational order) in $d=2$-$8$. While we cannot simulate dimensions high enough to show complete decorrelation, we show a loss of higher order structure that is consistent with both the decorrelation principle~\cite{torquato_new_2006} and the mean-field theory of hard-sphere liquids and glasses~\cite{parisi_meanfield_2010, parisi_theory_2020}. By understanding how structure diminishes in stealthy hyperuniform systems as a function of dimension, we can then begin to understand the role of spatial structure in determining the many novel properties of stealthy hyperuniform systems in the physical dimensions $d=1$-$3$~\cite{man_isotropic_2013, leseur_highdensity_2016, zhang_transport_2016, wu_effective_2017, froufe-perez_band_2017, gkantzounis_hyperuniform_2017, chen_designing_2018, torquato_multifunctional_2018, gorsky_engineered_2019, romero-garcia_stealth_2019, rohfritsch_impact_2020, sheremet_absorption_2020, kim_multifunctional_2020, romero-garcia_wave_2021, yu_engineered_2021, cheron_wave_2022,  klatt_wave_2022,  tavakoli_65_2022, granchi_nearfield_2022, kim_effective_2023, morse_generating_2023, kim_extraordinary_2024, kim_theoretical_2024}.

The goals of this work are then threefold. First, in Sec.~\ref{sec:defs} we give basic definitions and derive the constraints on all stealthy hyperuniform point patterns---both ordered and disordered---imposed by the hard-sphere condition. In Sec.~\ref{sec:densest}, we use these constraints to confirm that the densest Fourier-space configuration (i.e., the highest value of $K$ for a fixed number density of points) is a Bravais lattice, as stated in Ref.~\cite{torquato_ensemble_2015}. Then in Sec.~\ref{sec:disordered}, we discuss disordered stealthy hyperuniform systems and derive their effective Fourier-space packing fraction from the virial theorem. This then allows us to compare to numerical results showing decorrelation in Sec.~\ref{sec:decorrelation}. Finally, in Sec.~\ref{sec:conclusions}, we summarize our results and discuss the implications that this has on glasses.

\section{Definitions}
\label{sec:defs}
A Bravais lattice $\Lambda \in \mathbb{R}^d$ is a subgroup consisting of integer linear combinations of vectors constituting a basis for $\mathbb{R}^d$. There are $d$ basis vectors of $\Lambda$ labelled $\mathbf{a}_i$, and every point on the lattice can be specified as 
\begin{equation}
\mathbf{p}=\sum_{i=1}^d n_i \mathbf{a}_i
\end{equation}
where $n_i\in\mathbb{Z}$. A Bravais lattice is often simply referred to as a lattice in mathematics, so to avoid confusion, we will explicitly state whether a concept applies to Bravais lattices, non-Bravais lattices, or both. For Bravais lattices, space can be divided into identical regions $F$ called fundamental cells that each contain a single point $\mathbf{p}$. The volume of $F$ is denoted $v_F$. 

Every Bravais lattice has a dual Bravais lattice $\Lambda^*$, whose points are defined such that that if $\mathbf{p} \in \Lambda$ and $\mathbf{q} \in \Lambda^*$, then $\mathbf{p} \cdot \mathbf{q} = 2\pi m$ where $m\in\mathbb{Z}$. The dual fundamental cell $F^*$ has a volume $v_{F^*} = (2\pi)^d/v_F$
 such that the number density of the lattice $\rho_\Lambda$ and its dual $\rho_{\Lambda^*}$ are related as
 \begin{equation}
 \rho_\Lambda \rho_{\Lambda^*} = \frac{1}{(2\pi)^d}.
 \label{eq:latticeAndReciprocal}
 \end{equation}
While the concept of a dual lattice can be formally extended in some cases to periodic point patterns~\cite{cohn_ground_2009, cohn_formal_2014} (i.e., non-Bravais lattices), most periodic point patterns do not have a formal dual, and Eq.~\eqref{eq:latticeAndReciprocal} cannot in general be extended to them. 

In addition to lattices and explicit periodic point patterns, we consider disordered point patterns consisting of $N$ point particles in $\mathbb{R}^d$ embedded in periodic boxes $F$ with volume $v_F$ with positions $\mathbf{r}_j$. The structure factor for any of the aforementioned individual systems is given by
\begin{equation}
S(\mathbf{k}) = \frac{|\tilde{n}(\mathbf{k})|^2}{N}
\end{equation}
where $\mathbf{k}$ is a non-zero reciprocal lattice vector of $F$ and $\tilde{n}(\mathbf{k})$ is the complex collective coordinate of the wavevector $\mathbf{k}$ given by
\begin{equation}
\tilde{n}(\mathbf{k}) = \sum_{j=1}^N \exp(-i \mathbf{k}\cdot\mathbf{r}_j).
\end{equation}

Disordered stealthy hyperuniform systems can be obtained through the collective coordinate procedure~\cite{uche_constraints_2004, uche_collective_2006, batten_classical_2008, zhang_ground_2015, zhang_ground_2015a, morse_generating_2023}, which searches for a global minimum (ground state) of a potential related to $S(\mathbf{k})$ via numerical minimization. This potential energy can be derived as~\cite{zhang_ground_2015}:
\begin{equation}
\Phi(\mathbf{r}^N) = \frac{N}{2v_F}\bigg[\sum_{0<k\le K}\tilde{v}(\mathbf{k})S(\mathbf{k}) - \sum_{0<k\le K}\tilde{v}(\mathbf{k})\bigg].
\label{eq:energy}
\end{equation}
where $v(\mathbf{r})$ is a positive, bounded, and integrable function with compact support over the interval $0 < k \le K$ whose Fourier transform $\tilde{v}(\mathbf{k})$ exists. 
In practice, because the second term is a potential-dependent constant independent of the positions of the particles, it is dropped, and a disordered stealthy hyperuniofrm point pattern is then given by $\Phi(\mathbf{r}^N)=0$. While numerical simulations cannot reach this strict bound, they are able to achieve values of $\Phi$ which are indistinguishable from zero~\cite{finiteSizeSmax} to within double~\cite{uche_constraints_2004, uche_collective_2006, batten_classical_2008, zhang_ground_2015} or double-double precision~\cite{morse_generating_2023}.

While stealthy hyperuniform systems are characterized by the linear size of the exclusion region $K$, it is useful to consider the number of independently constrained wavevectors for which $S(\mathbf{k})=0$, labelled $M$. The total number of independent degrees of freedom in a system of $N$ particles is $(N-1)d$, and thus the fraction of independent degrees of freedom that are constrained, labelled $\chi$, is
\begin{equation}
\chi \equiv \frac{M}{(N-1)d}.
\label{eq:chiDef}
\end{equation}
The maximum value of $\chi$ for any disordered system is $\chi=\frac{1}{2}$ for $d\geq 2$~\cite{halfFootnote, torquato_ensemble_2015} (note also that $d=1$ is a special case---that we do not treat here---for which only $\chi<\frac{1}{3}$ are fully disordered; see Refs.~\cite{fan_constraints_1991}. All systems with $\chi>\frac{1}{2}$ are therefore ordered. In addition to crystalline states, the ordered regime contains stacked slider phases~\cite{uche_constraints_2004, zhang_ground_2015a}. Stacked sliders have implicit constraints that necessitate order (like a crystal), but these states do not contain Bragg peaks and are not periodic in real-space (see Refs.~\cite{uche_constraints_2004, zhang_ground_2015a} for visual examples).

Following the arguments of Ref.~\cite{torquato_ensemble_2015}, $\chi$ can be shown to act as an effective packing fraction of spheres with radius $K$ in Fourier-space by noting its value in the thermodynamic limit:
\begin{equation}
\chi = \frac{v_dK^d}{2d\rho(2\pi)^d},
\label{eq:chiThermo}
\end{equation}
where $v_d$ is the volume of a $d$-dimensional sphere of unit radius
\begin{equation}
v_d = \frac{\pi^{d/2}}{\Gamma(1+\frac{d}{2})}.
\end{equation}

\section{Densest Stealthy Hyperuniform Configurations}
\label{sec:densest}
There are two natural points of interest when discussing $\chi$ as an effective packing fraction: 1) the transition point between disordered and ordered systems at $\chi=\frac{1}{2}$ for $d\ge2$ and 2) the maximum obtainable value of $\chi$ in each dimension, which we denote $\chi_\mathrm{max}$. In Ref.~\cite{torquato_ensemble_2015}, it was shown that $\chi_\mathrm{max}$ is associated with the Bravais lattice that is the dual to the densest Bravais lattice in real-space. This section is devoted to finding an expression for $\chi_\mathrm{max}$ that makes this statement more explicit and that allows the full phase diagram of Fourier-space hard-spheres to be written in any dimension whose densest real-space lattice is known.

\begin{figure}[htp!]
\includegraphics[width=\columnwidth]{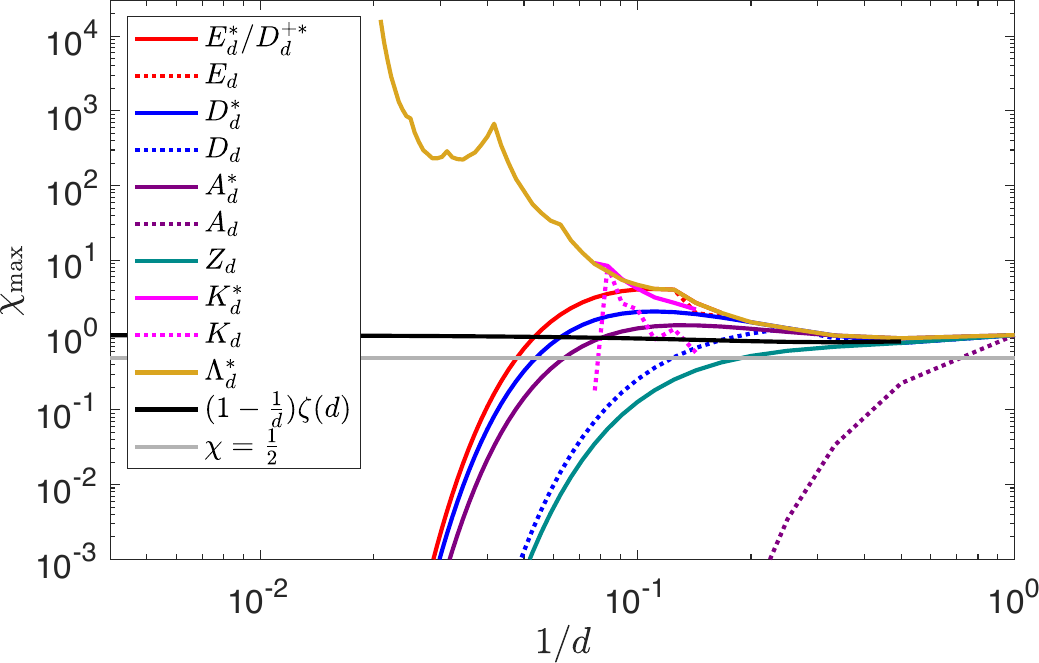}
\caption{Plot showing $\chi$ for root lattices, their duals, $K_{d}$ and its dual in $d=7$-$13$, and the duals of the laminated lattices $\Lambda_d$ in $d=1$-$48$. We also show the upper bound for disordered systems, $\chi=\frac{1}{2}$ (gray line), and the lower bound for the densest lattice in any given dimension $\chi=(1-\frac{1}{d})\zeta(d)$ (black line)~\cite{ball_lower_1992} showing that the densest Fourier-space hard-sphere packing in all dimensions is ordered.}
\label{fig:chimax_duals}
\end{figure}

To find the maximum value of $\chi$ for a given dimension, we first consider the maximum value of $\chi$ for a given Bravais lattice $\Lambda$, labelled $\chi_{\mathrm{max},\Lambda}$. Here, we can relate the number density of the reciprocal lattice to its packing fraction
\begin{equation}
\rho_{\Lambda^*} = \frac{2^d\varphi_{\Lambda^*}}{v_dK^d},
\end{equation}
which---combined with Eq.~\eqref{eq:chiThermo}---yields 
\begin{equation}
\chi_{\Lambda} = \frac{2^d \varphi_{\Lambda^*}}{2d} = \frac{\widehat{\varphi}_{\Lambda^*}}{2}.
\label{eq:chimaxLambda}
\end{equation}
Thus, the numerical value of $\chi$ associated with a given lattice is directly related to the packing fraction of its dual lattice in real-space. Values of $\chi_\Lambda$ for several standard lattices and their duals in $d=1$-$48$ are shown in  Fig.~\ref{fig:chimax_duals}. Notably, Eq.~\eqref{eq:chimaxLambda} makes explicit use of the scaling used to compare disorderd real-space hard-spheres across dimensions~\cite{torquato_controlling_2002, skoge_packing_2006, torquato_new_2006, parisi_meanfield_2010, parisi_theory_2020}
\begin{equation}
\widehat{\varphi} = \frac{2^d \varphi}{d}.
\label{eq:phiHat}
\end{equation} 
This scaling ensures that the minimal jamming packing fraction, the (avoided~\cite{charbonneau_glass_2017}) dynamical transition, and the onset of Fickian diffusion are all $\mathcal{O}(1)$ and smoothly evolve as $d\to\infty$~\cite{rscbook, mangeat_quantitative_2016, charbonneau_thermodynamic_2021, charbonneau_dimensional_2022, charbonneau_jamming_2023}. The appearance of this scaling when considering lattices is a subtle point, which has been noted elsewhere~\cite{ball_lower_1992, charbonneau_thermodynamic_2021, charbonneau_dimensional_2022}. Here, it will be used to make previous statements about the densest Fourier-space hard-sphere packing more apparent.

From Eq~\eqref{eq:chimaxLambda}, we define the absolute maximum $\chi_\mathrm{max}$ as the maximum value of $\chi_\Lambda$ taken over all lattices $\Lambda$ in a given dimension,
\begin{equation}
\chi_\mathrm{max} \equiv \max_\Lambda ( \chi_\Lambda).
\label{eq:chimax}
\end{equation}
We have two tools which allow for the calculation of $\chi_\Lambda$ when the maximum packing fraction of the dual lattice ($\varphi_{\Lambda^*}$) is unknown or when the dual lattice does not exist (i.e., in non-Bravais lattices). The first of these is an application of the stacked slider theorem~\cite{zhang_ground_2015a} 
\begin{equation}
\chi = \frac{v_dK_m^d}{2d(2\pi)^d\rho_P\rho_Q}
\end{equation}
where $P$ and $Q$ refer to lower dimensional lattices, for which $d_P + d_Q = d$ and $K_m = \min(K_P,K_Q)$. Applying this recursively allows the calculation of $\chi_\mathrm{max}$ in all laminated lattices. The second is an application of the Lemma on additive stealthy systems~\cite{torquato_ensemble_2015}, which states that for a system made additively from $m$ stealthy sub-systems, the combined value of $\chi$ is
\begin{equation}
\chi = \bigg[ \sum_{i=1}^m \chi_i^{-1} \bigg]^{-1}.
\end{equation}
Here, because $\chi_i > 0$ for all $i$, we have that $\chi < \chi_i$ for all $i$. Thus the value of $\chi$ for a crystal with an $m$ particle basis is lower than that of the Bravais lattice with the same unit cell. In Ref.~\cite{torquato_ensemble_2015} this was used to show that the densest lattice in Fourier-space is a Bravais lattice. From the construction of Eq.~\eqref{eq:chimaxLambda}, it is then clear that the  densest lattice in Fourier-space is the dual of the densest Bravais lattice in real-space---a statement noted in $d=1$-$4$ in Ref.~\cite{torquato_ensemble_2015}. This lattice was also proven to be the unique ground state in any given dimension~\cite{torquato_new_2008}.

Furthermore, it was shown~\cite{torquato_ensemble_2015} that the densest lattice in Fourier-space is more dense than the densest disordered system by logically extending the results of $d=1$-$4$. The explicit connection to real-space systems from Eq.~\eqref{eq:chimaxLambda} allows us to reinforce that proof. While the best lower~\cite{venkatesh_note_2013} and upper bounds~\cite{kabatiansky_bounds_1978, cohn_sphere_2014} on crystal close packing put the densest real-space Bravais lattice in the range
\begin{equation}
\max_\Lambda(\widehat{\varphi}_\Lambda) \in \bigg(65963, \frac{2^{0.401d}}{d}\bigg)
\end{equation}
as $d\to\infty$, a much weaker constructive lower bound is all that is necessary to prove that the highest $\chi_\mathrm{max}$ is associated with a Bravais lattice. For any given dimension, Ball showed that a Bravais lattice $\Lambda_B$ exists with ${\widehat{\varphi}_{\Lambda_B} \ge 2(1-\frac{1}{d})\zeta(d)}$ where $\zeta(d)$ is the Riemann zeta function~\cite{ball_lower_1992}, and thus
\begin{equation}
\chi_\mathrm{max} \ge \bigg(1-\frac{1}{d}\bigg)\zeta(d)
\end{equation}
for $d>1$. Note that for $d=1$ where $\zeta(1)$ diverges, $\chi_\mathrm{max}=1$. For positive integers $d$, $\zeta(d)\to1^+$ as $d\to\infty$, which converges rather quickly, meaning that $\chi_\mathrm{max}>\frac{1}{2}$ in all dimensions.

One caveat, which is clear from Fig.~\ref{fig:chimax_duals}, is that while $\chi_\mathrm{max}>\frac{1}{2}$ in all dimensions, individual Bravais lattices may have $\chi_\Lambda < \frac{1}{2}$ in any dimension, such as $Z_d$ for $d>5$. Additionally, crystals (i.e., lattices with a basis) may have $\chi_\Lambda < \frac{1}{2}$ even in low dimensions, such as kagom\'e ($\chi = 0.3022$) and honeycomb ($\chi=0.4534$) crystals in $d=2$ and the pyrochlore ($\chi=0.2267$) and diamond ($\chi=0.4534$) crystals in $d=3$~\cite{torquato_ensemble_2015}.

\section{Mapping to real-space disordered systems}
\label{sec:disordered}
While we have focused so far on the exclusion region $S(k<K)$, we would like to predict the behavior of the structure factor of disordered stealthy hyperuniform systems outside of the exclusion region, i.e., $S(k>K)$. In Ref.~\cite{torquato_ensemble_2015}, a subset of stealthy hyperuniform systems was identified for which this problem was tractable, namely, entropically favored states (EFS), which are first thermally equilibrated at a low temperature and then rapidly quenched through energy minimization. The EFS thus act like equilibrium Fourier-space hard-spheres. Following Ref.~\cite{torquato_ensemble_2015}, we use the mapping between a stealthy hyperuniform ensemble with $S(k)$ at an effective packing fraction $\chi$ and a real-space hard-sphere ensemble with $g_2^\mathrm{HS}(r)$ at real-space packing fraction $\widehat{\varphi}$ by
\begin{equation}
S(k,\chi) = g_2^\mathrm{HS}(r=k,\widehat{\varphi}),
\label{eq:hardSphereAnsatz}
\end{equation}
where $\chi = b(d)\widehat{\varphi}$. Because Eq.~\eqref{eq:hardSphereAnsatz} is only a statement of 2-point correlators, it should only be accurate for small values of $\chi$, though it should be increasingly accurate as $d$ increases and higher order correlations become trivial due to the decorrelation principle~\cite{torquato_new_2006}. Whether this mapping is precisely realizable in a given $d$ at any order of approximation is left as an open question. 

To first order in $\widehat{\varphi}$, the pair correlation function for equilibrated hard spheres can be written as~\cite{torquato_effect_2012, hansen_theory_2013}
\begin{equation}
g_2^\mathrm{HS}(r) = \Theta(r-\sigma)\big[1+\widehat{\varphi}\alpha(r;R)d+\mathcal{O}(\widehat{\varphi}^2)\big]
\end{equation}
where $\alpha(r;R)$ is the overlap of two spheres of radius $R$ separated by a distance $r$~\cite{torquato_effect_2012}
\begin{equation}
\alpha(r;R) = \frac{2\Gamma(1+\frac{d}{2})}{\pi^{1/2}\Gamma(\frac{d+1}{2})}\int_0^{\cos^{-1}[r/(2R)]}\sin^d(\theta) d\theta.
\end{equation}
The special case $r=R$, which obtains the maximum value of $\alpha(r;R)$, is then
\begin{equation}
\alpha(R;R) = \frac{{}_2F_1\big(\frac{1}{2},\frac{1+d}{2};\frac{3+d}{2};\frac{3}{4}\big)\Gamma(1+\frac{d}{2})\big(\frac{3}{4}\big)^{(1+d)/2}}{\pi^{1/2}\Gamma(\frac{3+d}{2})}
\end{equation}
where ${}_2F_1$ is the ordinary hypergeometric function. In infinite dimensions, $g_2^\mathrm{HS}(r)$ approaches a step function for all densities~\cite{frisch_classical_1985, torquato_new_2006, charbonneau_dimensional_2022}, which is reflected in the asymptotic form of the coefficient of $\widehat{\varphi}$ at the maximum value $r=R$:
\begin{equation}
\alpha(R;R)d \sim \sqrt{\frac{6d}{\pi}}\bigg(\frac{3}{4}\bigg)^{d/2}.
\end{equation}
As expected, this asymptotes to zero for all values of $\widehat{\varphi}$ as ${d\to\infty}$, and $g_2^\mathrm{HS}(r)$ approaches a step function.

\begin{figure}[htp!]
\includegraphics[width=\columnwidth]{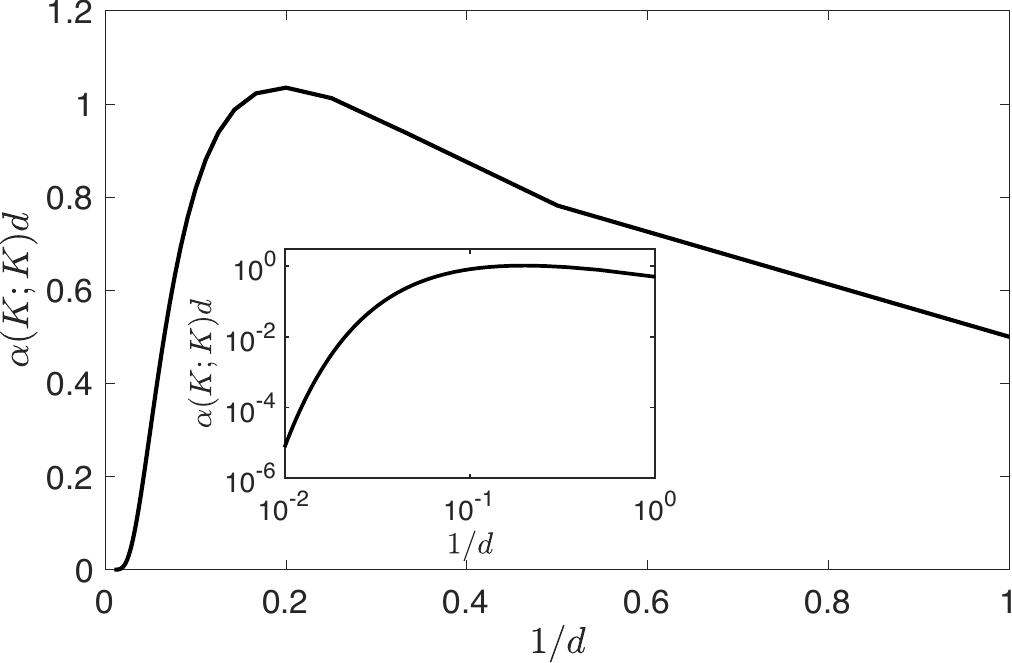}
\caption{The first order coefficient in the $\chi$ expansion of $S(k)$ for entropically favored states $\alpha(K;K)d$ as a function of dimension. We see that $\alpha(K;K)d \sim \mathcal{O}(1)$ for $d \lesssim 20$ but it then rapidly decreases as $d\to\infty$.}
\label{fig:alphad}
\end{figure}

Using the mapping of Eq.~\eqref{eq:hardSphereAnsatz}, we obtain~\cite{heavisideError}
\begin{equation}
S(k) = \Theta(k-K)\big[1+b(d)\chi\alpha(k;K)d +\mathcal{O}(\widehat{\varphi}^2)].
\end{equation}
In Ref.~\cite{torquato_ensemble_2015} it was argued that $b(d)=[\alpha(K;K)d]^{-1}$, noting the correct limit in $d=1$ and a strong fit to $d=2$ and $d=3$ data~\cite{bdunits}. However, this scaling gives an erroneous result in the prediction for $S(K)$ yielding $S(K) = 1 + \chi + \mathcal{O}(\chi^2)$. By construction, $S(k)$ should approach a step function as dimension increases, and thus $S(K)$ should approach $1$ as $d\to\infty$. In Sec.~\ref{sec:densest}, we noted that $\chi$ could be compared directly to real-space packing fractions in hatted units (see Eq.~\eqref{eq:phiHat}), suggesting that $b(d)=1$, and thus 
\begin{equation}
S(k) = \Theta(k-K)\big[1+\chi\alpha(k;K)d +\mathcal{O}(\chi^2)].
\label{eq:skEFS}
\end{equation}
Fig.~\ref{fig:alphad}, demonstrates that this achieves the appropriate limit to first order in $\chi$, because $\alpha(K;K)d\to0$ as $d\to\infty$. Over the densities considered in Ref.~\cite{torquato_ensemble_2015}, the differences between the choices of $b(d)=1$ and $b(d)=[\alpha(K;K)d]^{-1}$ yield curves which are almost indistinguishable, and thus Eq.~\eqref{eq:skEFS} fits EFS simulation data well.

It is, of course, important to note that the mapping of Eq.~\eqref{eq:hardSphereAnsatz} does not imply that all real-space hard-sphere properties can be directly transferred to Fourier-space hard-spheres. In Eq.~\eqref{eq:hardSphereAnsatz}, the mapping is only made at the two-particle level, which is most relevant at low densities and in high dimensions. We thus expect it to fail when considering highly correlated systems. It is thus unsurprising that the configuration with the maximum effective packing fraction in Fourier-space (Eq.~\eqref{eq:chimaxLambda}) is not what might have been expected using a straightforward substitution $\widehat{\varphi}\to\chi$. Similarly, the freezing point for Fourier hard spheres appears to be $\chi=\frac{1}{2}$, while for real-space hard spheres, $\widehat{\varphi}_f \sim 1$ in $d=3$-$10$ (see Ref.~\cite{charbonneau_thermodynamic_2021} for numerical values)~\cite{freezingFootnote}.

\begin{figure*}[htp!]
\includegraphics[width=\linewidth]{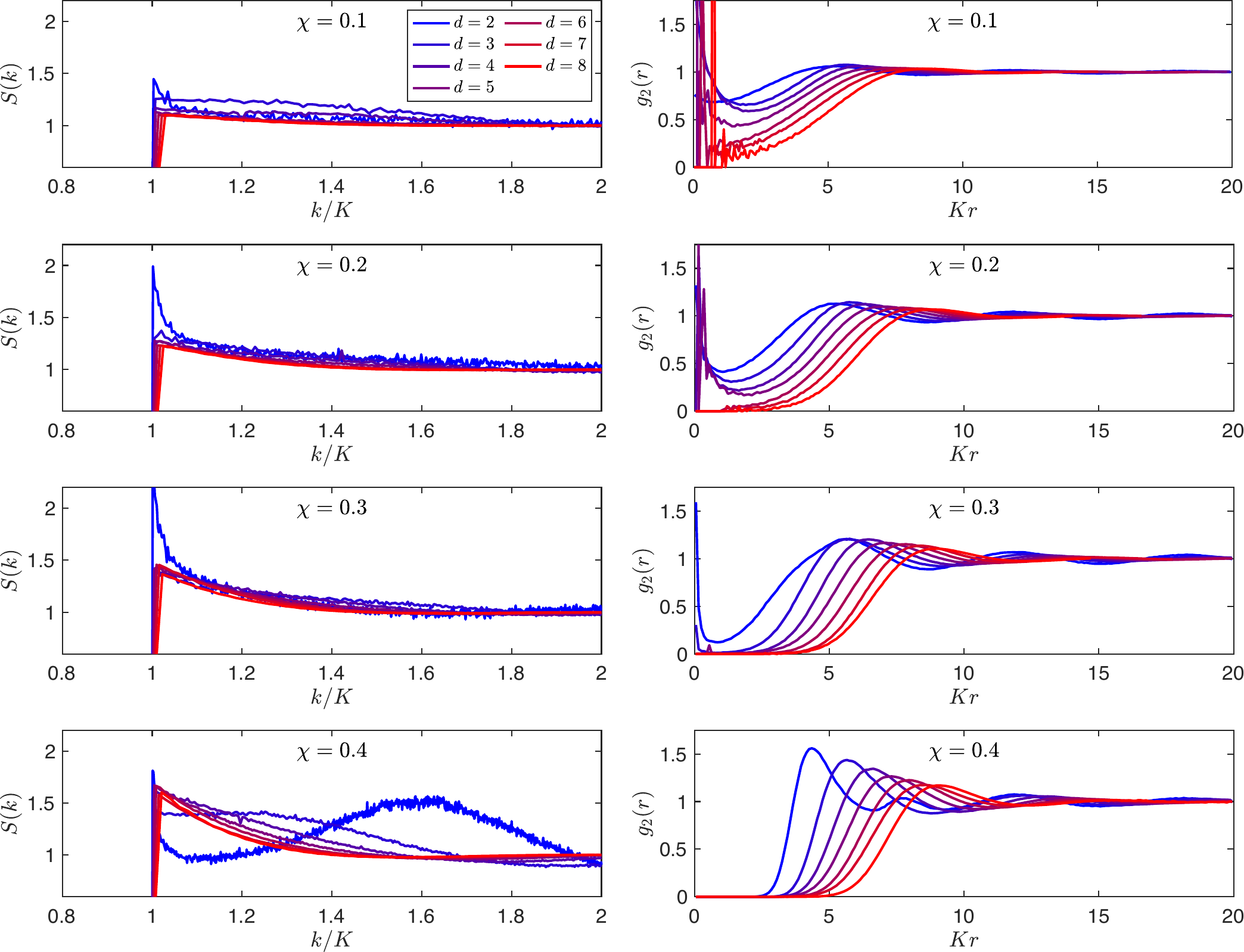}
\caption{(left) Plots of $S(k)$ (left) and $g_2(r)$ (right) at fixed $\chi$. As dimension increases, several prominent features in both metrics decrease, namely the peaks at low values of $k$ and $r$ respectively and oscillations in $S(k)$ at high values of $\chi$. As $d\to\infty$, $S(k)$ tends towards a step function for all values of $\chi$, however the dimensional range shown here is not enough for this trend to be clear, as the first correction to the step function in Eq.~\eqref{eq:skEFS} is $\mathcal{O}(1)$ for $d \lesssim 20$ (see Fig.~\ref{fig:alphad}). Instead, the curves appear to collapse towards a common curve as dimension increases. Noise in low $\chi$ high dimensional $g_2(r)$ data at small values of $r$ is due solely to low sampling.}
\label{fig:skgr}
\end{figure*}

\section{Evolution of structure in disordered systems}
\label{sec:decorrelation}
The collective coordinate minimization procedure described in Sec.~\ref{sec:defs} can be used to create disordered stealthy hyperuniform systems by minimizing $S(\mathbf{k})$ in the exclusion region. The behavior of the structure factor outside of the exclusion region $S(k>K)$ and the pair correlation function $g_2(\mathbf{r})$ will then depend largely on the initial conditions and the type of minimizer used~\cite{zhang_ground_2015}. While the theory of Ref.~\cite{torquato_ensemble_2015} gives theoretical predictions for both $g_2(r)$ and $S(k>K)$ in EFS, these states are difficult to obtain when $N$ is large, and the question of how to create EFS at large scales is left for future work. We instead focus on states with random initial conditions---equivalent to an infinite temperature quench (ITQ)---and ask whether key features predicted in EFS remain.

Fig.~\ref{fig:skgr} shows ITQ data in $d=2$-$8$ for $\chi=0.1$, $0.2$, $0.3$, and $0.4$. In $d=2$, we see large discrepancies between the theoretical values of $g_2(0)$ and $S(K)$ from EFS as already noted in Ref.~\cite{zhang_ground_2015} (see Fig. 1 therein) but these differences become negligibly small as dimension increases. For higher values of $\chi$, the non-monotonicity in $S(k)$ predicted in EFS are present in ITQ ensembles.
However, the oscillations in $S(k)$ at high $\chi$ and the spike near $k=K$ present in ITQ systems both decrease as dimension increases in agreement with the decorrelation principle~\cite{torquato_new_2006}.
We also note an apparent exclusion region forming in $g_2(r)$ for high values of $\chi$, which has been observed elsewhere in large but finite systems~\cite{uche_constraints_2004, zhang_ground_2015}
and exploited to map to packings of nonoverlapping
spheres.~\cite{zhang_transport_2016}. 

\begin{figure}[htp!]
\includegraphics[width=\columnwidth]{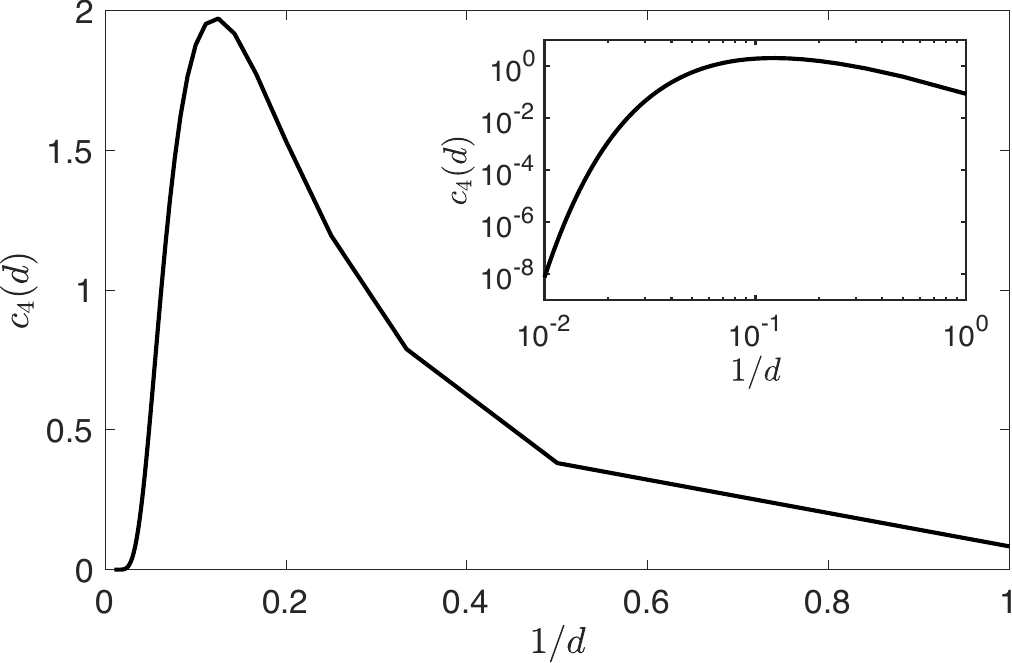}
\caption{The coefficient $c_4(d)$ of the $\chi^4$ term in the expansion of $\widehat{\tau}$ in Eq.~\eqref{eq:tauEFS} is plotted as a function of dimension. While $c_4(d)\sim\mathcal{O}(1)$ for $d \lesssim 20$, it then decreases rapidly as $d\to\infty$.}
\label{fig:c4}
\end{figure}

In addition to $g_2(r)$ and $S(k)$, we measure the degree of translational order through the order metric $\tau$~\cite{torquato_ensemble_2015}, given by
\begin{align}
\tau &\equiv \frac{1}{D^d}\int_{\mathbb{R}^d}\big[g_2(\mathbf{r})-1\big]^2d\mathbf{r} \nonumber \\ &= \frac{1}{(2\pi D)^d\rho^2}\int_{\mathbb{R}^d}\big[\langle S(\mathbf{k})\rangle-1\big]^2d\mathbf{k},
\end{align}
where angular brackets $\langle\dots\rangle$ denote an average over configurations and $D$ is a characteristic length, which we take to be $D=K^{-1}$. In an ideal gas $\tau = 0$, and thus $\tau$ represents the degree of departure from the fully uncorrelated case. To compare systems across dimension, we define a modified version, labelled $\widehat{\tau}$, which for isotropic systems ($g_2(\mathbf{r})=g_2(r)$) is
\begin{align}
\widehat{\tau} \equiv \frac{v_d\tau}{4d^2(2\pi)^d} &= \frac{K^dv_d^2}{4d(2\pi)^d}\int_0^\infty\big[g_2(r)-1\big]^2r^{d-1}dr \nonumber \\
&= \frac{K^dv_d^2}{4d\rho^2(2\pi)^{2d}}\int_0^\infty \big[\langle S(k)\rangle - 1 \big]^2k^{d-1}dk.
\end{align}
For EFS, to first order in $\chi$, this integral can be broken up into three regions using the form of Eq.~\eqref{eq:skEFS}
\begin{equation}
[S(k)-1]^2 = \begin{cases}
        1 & \text{if } k < K\\
        \chi^2\alpha(k;K)^2d^2 & \text{if } K\leq k \leq 2K\\
        0 & \text{if } k>2K.
    \end{cases}
\end{equation}
The integral over the region $k<K$ is trivially $\chi^2$, and the integral over the region $k>2K$ is trivially zero. The integral over the region $K\leq k \leq 2K$ can then be reduced to a nondimensional form to become 
\begin{align}
\frac{K^dv_d^2\chi^2d}{4\rho^2(2\pi)^{2d}}\int_K^{2K} \alpha(k;K)^2k^{d-1}dk &= \chi^4d^3 \int_1^2\alpha(t,1)^2t^{d-1}dt \nonumber \\
&= c_4(d)\chi^4,
\end{align}
where $c_4(d)$ can be calculated numerically. We can thus write 
\begin{equation}
\widehat{\tau} = \chi^2 + c_4(d)\chi^4 + \mathcal{O}(\chi^5).
\label{eq:tauEFS}
\end{equation}
Values of $c_4(d)$ rapidly decrease with dimension above $d\approx20$ (see Fig.~\ref{fig:c4}), meaning that $\widehat{\tau} \to \chi^2$ as $d\to\infty$, which is the result one would get  using $S(k) = \Theta(k-K)$. Given that the $\chi^2$ contribution will be common to all stealthy systems, we examine the \textit{excess} contribution to $\widehat{\tau}$ as $\widehat{\tau}-\chi^2$.

\begin{figure}[htp!]
\includegraphics[width=\columnwidth]{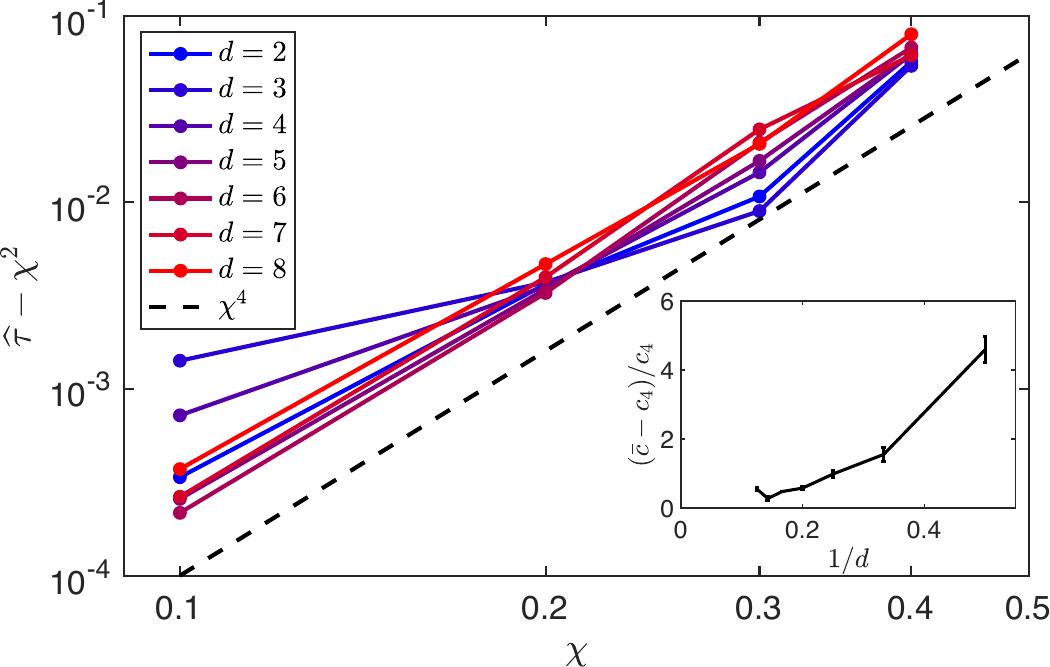}
\caption{The excess $\widehat{\tau}$ order metric $\widehat{\tau}-\chi^2$ as a function of both $d$ and $\chi$ from the $N=2\times10^6$ infinite temperature quench (ITQ) data (circles). For $d>4$, the ITQ data matches the functional form of the entropically favored state (EFS) prediction $\widehat{\tau}-\chi^2 \approx \bar{c}\chi^4$ (dashed black line). Lines between points are kept as a guide to the eye. The prefactor to the quartic term $\bar{c}$ also approaches $c_4(d)$ as $d$ increases, as evidenced in the inset, which shows the relative error of $\bar{c}$.}
\label{fig:tau}
\end{figure}

This can then be compared to the ITQ data (Fig.~\ref{fig:tau}), where we find remarkable functional agreement with the EFS prediction as $d$ increases, yielding $\widehat{\tau} - \chi^2 \approx \bar{c}\chi^4$ for $d>4$ with $\bar{c}$ approaching $c_4(d)$ as dimension increases. As noted in $d=2$~\cite{morse_generating_2023}, this indicates that the integrated measure of order across length scales is essentially the same between EFS and ITQ states, despite significant departures in both $S(k)$ and $g_2(r)$, and this is increasingly true as dimension increases.

The decorrelation effects seen in both $S(k)$ and $\tau$ indicate that disordered systems are also generally easier to find as dimension increases. This is borne out through the collective coordinate minimization procedure, which becomes significantly faster in higher dimensions, and which always finds a global minimum. This also strongly indicates that ground states will continue to exist for all $\chi<\frac{1}{2}$ as $d\to\infty$.

\section{Conclusions}
\label{sec:conclusions}
By elaborating on the analogy of stealthy hyperuniform points patterns as hard-spheres in Fourier-space, we have made several connections that allow insight into real-space packing problems. While the full phase diagram for real-space hard spheres is contested in high dimensions~\cite{charbonneau_three_2021}, the case for Fourier hard spheres is far simpler. In all dimensions $d\ge 2$, systems with $\chi < \frac{1}{2}$ are generically disordered ground states, with crystalline and stacked slider phases comprising a zero-measure set. For $\chi>\frac{1}{2}$, only crystalline and stacked slider phases are allowed, and we have derived a set of relations which clarify that the densest Fourier hard-sphere packing in any dimension is the dual to the densest Bravais lattice in real-space, as previously shown~\cite{torquato_ensemble_2015}. In $d=1$-$8$, the densest known Fourier-space hard-sphere packing happens to correspond with the best known solution to the quantizer problem~\cite{torquato_reformulation_2010}, though this correspondence is notably broken in $d=9$ and $10$~\cite{agrell_optimization_1998}.

Furthermore, features in $S(k>K)$, $g_2(r)$, and the $\tau$ order metric of disordered stealthy hyperuniform states decorrelate as dimension increases, implying that higher order correlations become trivial. This implies that disordered stealthy hyperuniform systems should be reconcilable with the infinite dimensional mean-field theory of glasses~\cite{parisi_meanfield_2010, parisi_theory_2020}, wherein all pair interactions between spheres are completely decorrelated. In infinite dimensions, the EFS become the full Fourier-space equivalent of real-space equilibrium hard spheres, whose dynamics and phase behavior are well understood within the mean-field theory. A full investigation of this mapping (and the existence or non-existence of a Gardner phase~\cite{gardner_spin_1985} in stealthy hyperuniform systems) is therefore within reach. 

\begin{acknowledgments}
The Research was sponsored by the Army Research Office and was accomplished under Cooperative Agreement Number W911NF-22-2-0103. Simulations were performed on computational resources managed and supported by the Princeton Institute for Computational Science and Engineering (PICSciE).
\end{acknowledgments}

\bibliographystyle{apsrev4-1}
\bibliography{hyperuniform,footnotes}

\begin{thebibliography}{69}
\bibitem [{fin()}]{finiteSizeSmax}%
  \BibitemOpen
  \bibinfo {note} {The precise value that $\Phi$ acheives will depend on $N$,
  $d$, $K$, and $\tilde{v}(\mathbf{k})$.}\BibitemShut {Stop}%
\bibitem [{hal()}]{halfFootnote}%
  \BibitemOpen
  \bibinfo {note} {The value $\chi = \frac{1}{2}$ comes from the constraint
  that both the real and the imaginary components of $\tilde{n}(\mathbf{k})$
  must be zero in order for $S(\mathbf{k})=0$.}\BibitemShut {Stop}%
\end{thebibliography}

\end{document}